\documentclass[conference]{IEEEtran}
\IEEEoverridecommandlockouts
\usepackage{cite}
\usepackage{amsmath,amssymb,amsfonts}
\usepackage{amssymb}
\usepackage{algpseudocode}
\usepackage{amsfonts}
\usepackage{graphicx}
\usepackage{fancyhdr}  %
\usepackage{cases}
\usepackage{textcomp}
\usepackage{extarrows}
\usepackage{algorithm,algpseudocode}
\usepackage{multirow}

\usepackage{multirow,tabularx}
\usepackage{mathtools}
\usepackage{xcolor}
\usepackage[english]{babel}
\usepackage{amsthm}

\def\BibTeX{{\rm B\kern-.05em{\sc i\kern-.025em b}\kern-.08em
    T\kern-.1667em\lower.7ex\hbox{E}\kern-.125emX}}
\begin{document}

\title{Robust mmWave Beamforming by Self-Supervised Hybrid Deep Learning}

\author{Fenghao Zhu, Bohao Wang, Zhaohui Yang, Chongwen Huang, Zhaoyang Zhang,  \\ George~C.~Alexandropoulos, Chau~Yuen and M\'{e}rouane Debbah

\thanks{F. Zhu, B. Wang, Z. Yang, C. Huang and Z. Zhang are with the College of Information Science and Electronic Engineering, Zhejiang University, Hangzhou, 310007, China (e-mail:chongwenhuang@zju.edu.cn).}

\thanks{G.~C.~Alexandropoulos is with the Department of Informatics and Telecommunications, National and Kapodistrian University of Athens, Panepistimiopolis Ilissia, 15784 Athens, Greece (e-mail: alexandg@di.uoa.gr).}

\thanks{C. Yuen is with the School of Electrical and Electronics
Engineering, Nanyang Technological University, Singapore }

\thanks{M. Debbah is with Khalifa University of Science and Technology, P O Box 127788, Abu Dhabi, UAE (email: merouane.debbah@ku.ac.ae)}
%	\author{
%	\IEEEauthorblockN{Fenghao Zhu\IEEEauthorrefmark{1}, Bohao Wang\IEEEauthorrefmark{1}, Chongwen Huang\IEEEauthorrefmark{1} }
%	\IEEEauthorblockA{\IEEEauthorrefmark{1} Department of Information and Electronic Engineering, Zhejiang University}
}

\maketitle

\begin{abstract}
Beamforming with large-scale antenna arrays has been widely used in recent years, which is acknowledged as an important part in 5G and incoming 6G. Thus, various techniques are leveraged to improve its performance, e.g., deep learning, advanced optimization algorithms, etc. Although its performance in many previous research scenarios with deep learning is quite attractive, usually it drops rapidly when the environment or dataset is changed. Therefore, designing effective beamforming network with strong robustness is an open issue for the intelligent wireless communications. In this paper, we propose a robust beamforming self-supervised network, and verify it in two kinds of different datasets with various scenarios. Simulation results show that the proposed self-supervised network with hybrid learning performs well in both classic DeepMIMO and new WAIR-D dataset with the strong robustness under the various environments. Also, we present the principle to explain the rationality of this kind of hybrid learning, which is instructive to apply with more kinds of datasets.
\end{abstract}


\section{Introduction}\label{sec:intro}
Recently, beamforming for millimeter wave (mmWave) has gained much attention for its very wide bandwidth and the capacity of forming the narrow beam, which can greatly improve the throughput of communication systems. Narrow beam makes the directivity of beamforming better so that the energy would be concentrated and the signal to noise rate would be higher. Beamforming technologies had been thoroughly studied in the past decades \cite{beamformingsolution}. Specifically, exhaustive search and adaptive hierarchical search were first proposed to solve the problem, but at the price of the large training overhead and delay. Majorization Minimization algorithm is an efficient way to get the sub-optimization solution\cite{mmalgorithm}, however it usually does use accurate channel information, which will results in the heavy training overhead.
\par
Machine learning (ML) is good at exploiting historical experience to solve the new similar problems, and previous works had shown that ML-based technology can be used to predict mmWave beams\cite{sub6g}. Moreover, a recent work has presented an universal framework to Multiple Input Single Output (MISO) beamforming\cite{universalMISO}, but all of these works lack the ability to migrate to different scenes or even open environment. This is also the defect of the most ML-based beamforming methods. Therefore, implementing a robust beamforming framework is a very meaningful and important topic, especially when it can work well with different kinds of datasets. First, we will introduce these two kinds of different datasets. It is well known that DeepMIMO is the most widely used open-source dataset in wireless communications, which is constructed by the 3D ray-tracing and virtual scene generation technologies \cite{alkhateeb2019deepmimo}. Due to its integrity, it can simulate the most classical scenes. However, the shortcoming is also very obvious: it is based on the virtual generated scene, which might not provide accurate simulations for the real communications. Moreover, this dataset just provides one setting for each kind of scene, and the trained model can not adapt to various building settings.

Recently, another an open-source dataset named as WAIR-D was proposed to boost the related research\cite{WAIR-D}. The biggest difference from DeepMIMO is that it is based on the real-scenario maps. In addition, the dataset generating and saving ways  are different as well. The main features of WAIR-D can be summarized as follows:
\begin{itemize}
	\item \textit{Real Scenes:} 10000 scenes are randomly selected from real-scenario maps of more than 40 big cities around the globe, and an example can be seen from Fig$.$~\ref{fig:real_map}, where the layout information of buildings is directly extracted from the actual map.
	\item \textit{Flexible Parameters:} 3D ray-tracing parameters are provided so that the user can define them freely to generate required data  for wireless AI tasks.
	\item \textit{Friendly to Users:} Data genaration, preprocessing, sample, and training codes are all provided for users to get the start quickly.
     \item \textit{Friendly to Low-memory computers:} The number of threads for data generation can be adjusted so that less memory is taken to reduce the time overhead of reading and training.

\begin{figure}[h]\vspace{-2mm}
	\begin{center}
		\centerline{\includegraphics[width=0.42\textwidth]{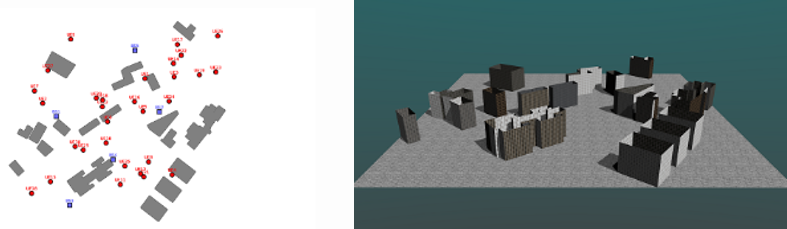}}  \vspace{-0mm}
		\caption{An example environment with buildings and BS(blue)/UE(red) positions in top view (left), the same buildings in bird-eye view (right)}
		\label{fig:real_map} \vspace{-2mm}
	\end{center}
\end{figure}
\vspace{-0mm}
\end{itemize}

In this work, we will present a robust beamforming self-supervised scheme based on these two kinds of different datasets, where no label is needed and the proposed beamforming network can be trained to learn how to optimize the spectral efficiency intelligently. Finally, we verify the robustness performance of our proposed scheme with various scenarios. The main contributions can be summarized as follows:
\begin{itemize}
	\item \textit{Robustness inside Dataset:} We design a self-supervised beamforming neutral network to produce analog phase shifters instead of digital ones, which is more consistent with mmWave. With different test cases, it can still output the optimized beamformer, which implies the neutral network that has robustness inside a specific dataset.
	\item \textit{Robustness across Dataset:} Hybrid pre-training is proposed to gain a more robust beamforming network to adapt to a different environments. Moreover, we give the principle to explain the rationality of this hybrid learning, which is instructive to apply with more kinds of datasets.
	\item \textit{Multiple Data Sources Performance Evaluation:} We verify our proposed scheme in two kinds of different datasets with various scenarios. Simulation results show that the proposed self-supervised network with hybrid learning performs well in both classic DeepMIMO and new WAIR-D dataset with the strong robustness under the various environments.
\end{itemize}

%The remainder of this paper is organized as follows. In Section \ref{sec:format}, the considered system model is given with the problem formulation. Details of the joint channel estimation and signal recovery scheme are provided in Section \ref{sec:channel_est}. Section~\ref{sec:simulation} presents the estimation performance under some considered scenarios. Finally, concluding remarks are drawn in Section~\ref{sec:conclusion}.

%\textit{Notation}: Fonts $a$, $\mathbf{a}$, and $\mathbf{A}$ represent scalars, vectors, and matrices, respectively. $\mathbf{A}^T$, $\mathbf{A}^H$, $\mathbf{A}^{-1}$, $\mathbf{A^\dag}$, and $\|\mathbf{A}\|_F$ denote transpose, Hermitian (conjugate transpose), inverse, pseudo-inverse, and Frobenius norm of $ \mathbf{A} $, respectively. $a_{mn}$ is the $(m,n)$-th entry of $\mathbf{A}$ and $|\cdot|$  denotes the modulus. $o$ represents infinitesimal of higher order. Finally, notation $diag(\mathbf{a})$ and $Tr(\mathbf{a})$ represent the diagonal matrix and trace of $\mathbf{a}$ respectively.

\section{System Model}\label{sec:format}
% \subsection{System Model}\label{subsec:signal model}
%\begin{figure}[b]\vspace{-0mm}
%	\begin{center}
%		\centerline{\includegraphics[width=0.42\textwidth]{system_model.png}}  \vspace{-0mm}
%		\caption{The considered MISO system consisting of a $N_t$-antenna base station serving a  %single-antenna mobile user.}
%		\label{fig:Estimation_Scheme} \vspace{-4mm}
%	\end{center}
%\end{figure}
%Due to the heavy attenuation of mmWave,
In this section, we first introduce the system model for the considered mmWave communication scenario, and then present a rate-maximization optimization problem. We consider the downlink MIMO-based mmWave communication system\cite{holographic}, where there is a base station (BS) equipped with an array of $N_t$ antenna elements, and $N_{RF} $ radio frequency chains. In the receiver,  there are $N_r$ single-antenna mobile users. Due to the heavy attenuation of mmWave, we only consider the line of sight (LOS) dominant paths between BS and users. We denote the transmitted symbol as $\mathbf{x} \in C^{N_{RF}} \times 1  $, with average energy being $E\left\{|\mathbf{x}\mathbf{x}^H|\right\}=I$. We adopt a narrowband block-fading channel model in which the $k\text{-}th$ user observes the received signal as
\begin{equation}\label{Transmission Model}
y =\mathbf{h}_k^H\mathbf{v}_k \mathbf{x} + n,
\end{equation}
where $\mathbf{h}$ is the $ N_t \times 1$ vector that represents the mmWave
channel between the BS and user, and $\mathbf{v} \in C^{N_t \times N_{RF} }$ is the beamforming matrix. $n$ is the additive noise satisfying the circularly symmetric complex Gaussian distribution with zero mean and covariance $\sigma^{2}$.

\begin{figure}[h]\vspace{-2mm}
	\begin{center}
		\centerline{\includegraphics[width=0.33\textwidth]{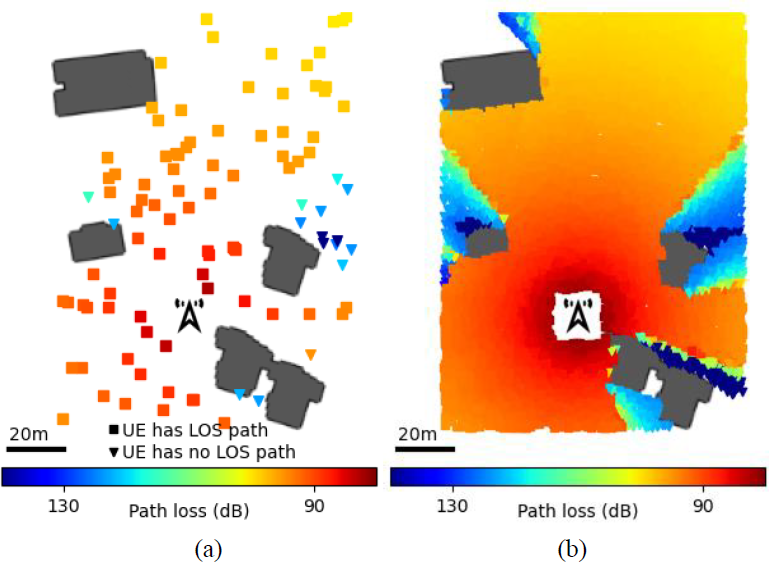}}  \vspace{-0mm}
		\caption{Top view of the scenario 2 case 1 of the WAIR-D.(a) shows 100 UEs among the totally 10,000 UEs, (b) shows all 10,000 UEs}
		\label{fig:wair-d_map} \vspace{-4mm}
	\end{center}
\end{figure}

As for the channel model, we now have generated two datasets: DeepMIMO and WAIR-D. The time-domain channel matrix of DeepMIMO, $\mathbf h_{d}$ consists of $L$ channel paths, and each path has a time delay $\tau_{l} \in R$, and the azimuth/elevation angles of arrival (AoA) $\theta_{l}, \varphi_{l}$. Let $\rho$ denote the path-loss between the user and the BS, and $p(\tau)$ represents a pulse shaping function for $T_{S}$-spaced signaling evaluated at $\tau$ seconds. Thus, the channel vector can be formulated as
\begin{equation}\label{DeepMIMO Channel Model}
\mathbf h_{d}= \sqrt \frac{M}{\rho}\sum_{l=1}^{L}\alpha_{l} p(dT_{S}-\tau_{l})  \mathbf a(\theta_{l}, \varphi_{l}),
\end{equation}
where $\mathbf a(\theta_{l}, \varphi_{l})$ is the array response vector of the BS at the AoA
$\theta_{l}, \varphi_{l}$. By contrast, WAIR-D has the same channel model but is constructed based on real map in large scale, which makes it more accurate to simulate the real scenes. The $k\text{-}th$ user can obtain the spectral efficiency as follows \cite{largescale},
\begin{equation}\label{Spectral Efficiency}
R_k(\mathbf v_k)= \log (1+\frac{\|\mathbf {h}_k^{H}\mathbf{v}_k\|^{2}}{\sigma^{2}+\sum_{j \neq k }\|\mathbf {h}_k^{H}\mathbf{v}_j\|^{2}}).
\end{equation}
where the beamforming vector satisfies power constraint $\sum_k\|\mathbf{v}_k^H\mathbf{v}_k\| \leq P$. And signal-noise ratio (SNR) can be defined as $10\lg(P/\sigma^{2})$. Since this task can be formulated as a mapping $\mathbf v_k=\mathcal V(\mathbf h_k, P)$, the optimization problem can be written as
\begin{equation}\label{Target Spectral Efficiency}
\mathop {\rm max}\limits_{ \mathbf v_k}  \mathbb{E}_{\mathbf h,P}[\sum_k  R_k( \mathcal V(\mathbf h_k, P))]
\end{equation}
\begin{equation}\label{constraint}
	s.t. \sum_k\|\mathcal V(\mathbf h_k, P)^H \mathcal V(\mathbf h_k, P)\| \leq P,
\end{equation}
where $\mathbb{E}_{\mathbf h,P}[ \cdot ]$ represents average over channel samples and the sum power constraints. The main idea of this paper is to focus on how the spectral efficiency changes with SNR, and channel state information (CSI) is what we will utilize to predict the best beamforming vector, although the estimation of it is imperfect. Accuracy of channel estimation can be measured by pilot-to-noise power ratio (PNR), moreover, \cite{largescale} had shown that deep neural network can approximate theoretical upper bound in high PNR. To verify the network robustness across different datasets while reducing impact of estimation bias, we set the appropriate PNR as the 20dB.
\begin{figure}[h]\vspace{-2mm}
	\begin{center}
		\centerline{\includegraphics[width=0.35\textwidth]{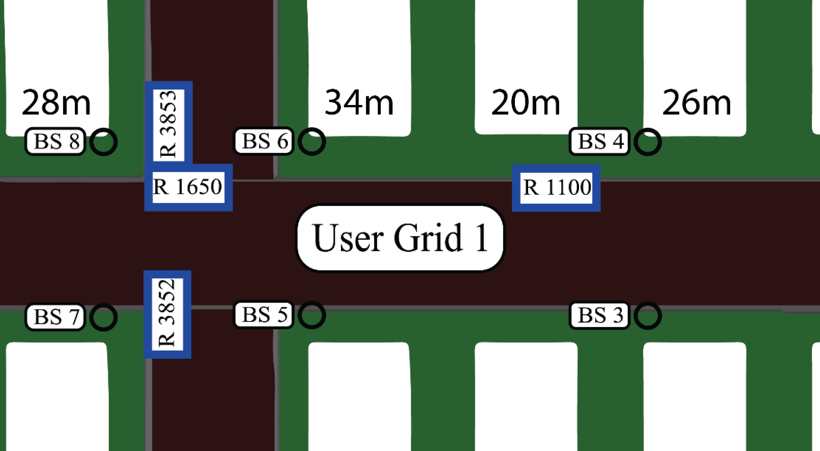}}  \vspace{-0mm}
		\caption{Top view of the scenario 'O1' of the DeepMIMO dataset.}
		\label{fig:deepmimo_map} \vspace{-4mm}
	\end{center}
\end{figure}
The scenario that we choose in DeepMIMO is O1\_60. In terms of WAIR-D dataset, scenario-1 of WAIR-D has 30 users and 4 stations, where the user number is not enough, therefore, we choose scenario-2 in  WAIR-D for our experiment, and details of the training maps are illustrated in Fig$.$~\ref{fig:deepmimo_map} and Fig$.$~\ref{fig:wair-d_map} respectively.

Related research\cite{blockage} has pointed out that multi-link interference in such beamforming task is limited, and taking interference into consideration would require a more complex network architecture and much higher computation complexity. Moreover, the impact of multi-link interference on the performance of prediction is expected to be small due to mmWave beamforming itself\cite{attenuation}. So our simulation would not consider interference between users, although our model is a general one applying to multi users.
%are expressed as follows:
%\begin{equation}\label{mode_1}
%\text{Mode-1:} \quad {\mathbf{Z}_1=({\mathbf{H}^e}^T \circ   \mathbf{\Phi}){\mathbf{H}^r}^T}\in\mathbb{C}^{PT \times K},
%\end{equation}
%\begin{equation}\label{mode_2}
%\text{Mode-2:}  \quad {\mathbf{Z}_2=( \mathbf{\Phi} \circ \mathbf{H}^r)\mathbf{H}^e}\in\mathbb{C}^{KP \times T},
%\end{equation}
%\begin{equation}\label{mode_3}
%\quad \text{Mode-3:} \quad {\mathbf{Z}_3=( \mathbf{H}^r \circ  {\mathbf{H}^e}^T)  \mathbf{\Phi}^T}\in\mathbb{C}^{TK \times P}.
%\end{equation}

%In the following section,  we present the joint estimation and recovery approach based on these unfolded forms.
\vspace{-1mm}

\section{Proposed Self-Supervised Beamformer }\label{sec:channel_est}
\begin{figure}[h]\vspace{-1mm}
	\begin{center}
		\centerline{\includegraphics[width=0.52\textwidth]{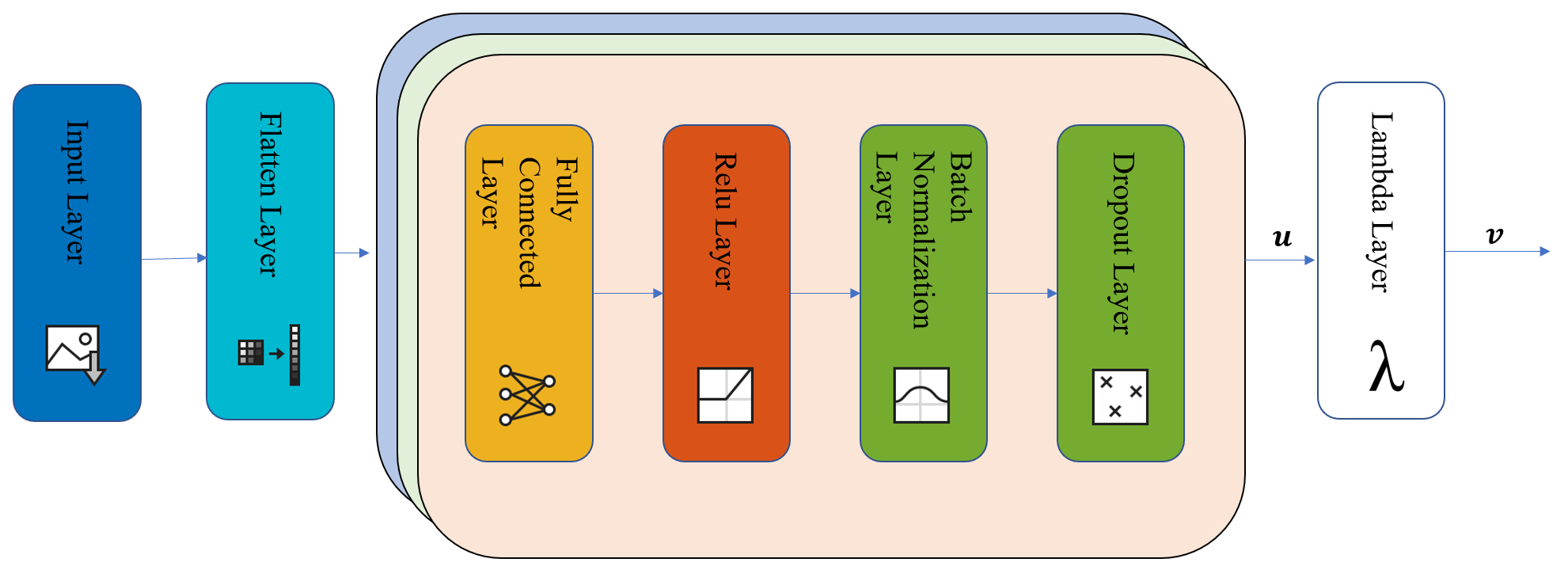} } \vspace{-2mm}
		\caption{The architecture of beamforming network.}
		\label{fig:framework}
      \vspace{-6mm}
	\end{center}
\end{figure}

In this section, we present the unsupervised machine learning architecture, and discuss how to combine datasets to improve hybrid training efficiency in the proposed network architecture, which is shown in Fig$.$~\ref{fig:framework}, and it is mainly composed of three parts: input block, feature extracting block and output block. Such design can capture relationship between layers and effectively alleviate overfitting, making no special assumptions about input data. Afterwards, a hybrid training theory is presented to deduce the robustness, making our model work around different environments. The details are presented in following subsections.

  \vspace{-2mm}

\subsection{The Network Architecture}
CSI of WAIR-D usually can be provided as an image form to save the storage space, the $x$-axis and $y$-axis of which represent number of carriers and antennas separately, and the deepth of color represents channel frequency response (CFR). CSI of DeepMIMO is provided as a complex matrix, and we can split the real part and negative part and concatenate them to form an image. Thus, we can use image as the form of input. After flatten layer, input data becomes one-dimension and is ready to be processed by the next layer. The second block mainly consists of three smaller blocks, in which fully connected layer is mainly used to extract features, and it is followed by relu layer, batch normalization layer and dropout layer. Batch normalization layer can normalize the biased distribution to standard distribution. In this way, the gradient becomes larger; the learning convergence speed is accelerated, and the problem of gradient disappearance is alleviated. And related research had shown that batch normalization layer should be put between relu layer and dropout layer\cite{batchnormalization}, so we design this sequence to maximize robustness. The reason why we choose three smaller blocks is that through experiment we have found that this is the best balance between complexity and performance. The detailed parameters in our simulation can be found in table \ref{network}.
\begin{table}[h] \vspace{-2mm}
\centering
\caption{DETAILS OF THE NETWORK}
\label{network}
\begin{tabular}{||c c c||}
 \hline
 Layer Name & Output Size & Parameters  \\ [1.0ex]
 \hline\hline
 Input Layer & $2 \times 64$  & 0 \\
 Dense Layer 1 & $ 320 \times 1$ & 41280  \\
 BN Layer 1 & $ 320 \times 1$ & 640  \\
 Dense Layer 2 & $ 320 \times 1$ & 102720 \\
 BN Layer 2 & $ 320 \times 1$ & 640  \\
 Dense Layer 3 & $ 128 \times 1$ & 41088 \\
 BN Layer 3 & $ 128 \times 1$ & 256  \\
 Lambda Layer & $ 64 \times 1$ & 0 \\ [1.0ex]
 \hline
\end{tabular}
\end{table}\vspace{-0mm} 
\par
The last dropout layer outputs $2N_tN_rN_{RF}$ real scalars, corresponding to $N_r$ complex vectors, and $\mathbf u = [\mathbf u_1, \mathbf u_2 \cdots, \mathbf u_{N_r}]$ is the combination. Constrained by $\sum_k\|\mathbf{v}_k^H\mathbf{v}_k\| \leq P$, $\lambda$ layer utilizes generated complex vector $\mathbf u$ to output the final beamforming vector $\mathbf v = [\mathbf v_1, \mathbf v_2 \cdots, \mathbf v_{N_r}]$ as
\begin{equation}\label{Final Beamforming Vector}
\mathbf v_k= \sqrt \frac{P}{\sum\limits_{ 1 \leq j \leq N_r}\| \mathbf u_j\|^{2}}\mathbf u_k.
\end{equation}
As the feature of self-supervised learning, we do not need labels before our training process, where we only need CSI and power $P$. The training loss is the opposite number of the average transmission rate, as we directly output the beamforming vector and transmission rate. Stochastic gradient descent method (SGD) is utilized to approach the max rate and the loss function takes the form
\begin{equation}\label{Loss Function}
\rm{Loss} =- \frac{1}{N_r M}\sum_{i=1}^{M}\sum_{k=1}^{N_r} R_{k,i}(\mathbf v_k).
\end{equation}
$M$ is the size of training batch while $R_{k,i}$ refers to the spectral efficiency of the $k\text{-}th$ user in the $i\text{-}th$ sample, and each sample contains channel data of $N_r$ users.

\subsection{Hybrid Training Theory}
In this section, we propose a hybrid training method for beamforming and present a theory to explain it. We assume that the training data is combined of $K$ kinds of datasets with $n$ samples in total, and the proportion of each dataset is $q_{k}$ and $p_{k}$ is the according CSI of an original specific dataset. It is clear that $\sum_{k=1}^{K}q_{k}=1$ and amount of channel vectors of dataset $k$ is $nq_k$. Thus, this mixed dataset can be defined as $p=\sum_{k=1}^{K}q_{k}p_{k}$, and $\mathbf q =(q_1,q_2\cdots,q_K)$. We assume that the parameter of neural work is $\mathbf \Theta$; input CSI is $\mathbf h$, and the output spectral efficiency is $R$. $\ell(\mathbf \Theta;\mathbf h,R)$ is the train sample loss, which is used to assess the gap between the predicted value and the actual maximum value, although the actual optimal value is not required in self-supervised learning and can not be accessed directly. Therefore, the search of the optimal parameter can be expressed as
\begin{equation}\label{minloss}
\mathbf \Theta(n, \mathbf q)_{optimal}= \mathop{ \rm argmin}\limits_{ \mathbf \Theta}\mathbb{E}_{n,\mathbf q}[\ell(\mathbf \Theta;\mathbf h,R)].
\end{equation}
Since we are only interested in how $n$ and $\mathbf q$ impact the training loss, we can formulate a function to assess the extra averaged training loss,
\begin{equation}\label{extraloss}
L(n, \mathbf q) =\mathbb{E}_{n,\mathbf q}[\ell(\mathbf \Theta;\mathbf h,R)]-\left \{ \mathbb{E}_{n,\mathbf q}[\ell(\mathbf \Theta;\mathbf h,R)]\right \}_{min}.
\end{equation}
Earlier research had shown that there is a log-linear relationship between $L(n, \mathbf q)$ and $log(n)$, if $\mathbf q$ is fixed\cite{predictable}, so the extra loss can be represented by
\begin{equation}\label{extralinearloss}
\log(L(n, \mathbf q)) \approx \alpha(\mathbf q)\log(n)+C(\mathbf q).
\end{equation}
Further, $\alpha(\mathbf q)$ was found to be a constant in a specific task regardless of data composition\cite{iclr}, and this decouples $n$ and $\mathbf q$, helping us concentrate on the proportion parameter $\mathbf q$. When $n$ is fixed, $\alpha(\mathbf q)\log(n)$ turns into a constant, then the extra averaged training loss can just be described by $C(\mathbf q)$. At the same time a rational function was proposed\cite{icml} to fit $C(\mathbf q)$ written as
\begin{equation}\label{Cq}
C_{\lambda}(\mathbf q)=\sum_{i=1}^{d}\left(\sum_{k=1}^{K}\lambda_{ik}q_{k}\right)^{-1},
\end{equation}
in which $\lambda_{ik}$ is a parameter determined by this model and $d$ is the number of elements in channel vector $\mathbf h$. The formulation has been proved effectively by the empirical and theorical way so that we can focus on the impact of $\mathbf q$ on the training loss independently. But the process of  getting the parameter $\lambda_{ik}$ is not so straightforward, we first make an assumption that the loss $\ell$ is twice differentiable (true if the network can backpropagate), and $n$ is fixed and large enough, thus, we can define the two variables as below
\begin{equation}\label{sigmak}
\sum\nolimits_k =\mathbb{E}_{p_k}[\nabla \left(\ell(\mathbf \Theta;\mathbf h,R)\right)].
\end{equation}
\begin{equation}\label{sigmastar}
\sum\nolimits^* =\mathbb{E}_{p^*}[\nabla \left(\ell(\mathbf \Theta;\mathbf h,R)\right)].
\end{equation}
here $\nabla \left(\ell(\mathbf \Theta;\mathbf h,R)\right)$ referes to Hessian-Matrix of $\ell(\mathbf \Theta;\mathbf h,R)$ with elements of $\mathbf h$ being variables. And $p^*$ refers to CSI of test dataset mixed in a certain proportion.  So $C(\mathbf q)$ can be approximated\cite{icml} as
\begin{equation}\label{trcq}
C(\mathbf q) \approx Tr\left(\sum\nolimits^*\left(\sum_k\left(q_k\sum\nolimits_k\right)\right)^{-1}\right),
\end{equation}
wherein $Tr$ represents trace of matrix. It implies that $C(\mathbf q)$ is not only associated with the training dataset but also is associated with the validation dataset. If $\sum\nolimits^*$ can be diagonalized $P^{-1}\sum\nolimits^*P = D^*$ and $\sum\nolimits_k$ can also be approximately diagonalized $P^{-1}\sum\nolimits_k P= D_k+O_k$ with the same orthogonal matrix $P$ and small enough $O_k$, by substituting (\ref{sigmak}) and (\ref{sigmastar}) into (\ref{trcq}), we can find the parameter  $\lambda_{ik}$ by the following formula:
\begin{equation}\label{trlambda}
\begin{aligned}
C(\mathbf q) &\approx \sum\limits_{i=1}^{d}\frac{1}{\sum_{k=1}^{K}q_k\frac{D_{k,ii}}{D_{ii}^*}}+o\left(\| \sum_{k=1}^{K}q_kO_k \|_F\right).
\end{aligned}
\end{equation}
$D_{k,ii}$  refers to the $i\text{-}th$ element in the diagonal of $D_{k}$. Now we get $\lambda_{ik}$ in \eqref{Cq} which equals $\frac{D_{k,ii}}{D_{ii}^*}$, and this can be implemented through computing Hessian-Matrix and matrix diagonalization shown above. If the neural network has been trained with a mixed dataset of proportion $\mathbf q$, (15) can be derived using steps illustrated, therefore we can compare the actual performance of the model with $C(\mathbf q)$. It is to be noted that if the diagonalization requirement is not met, $C(\mathbf q)$ is still accessible through computing Hessian-Matrix to get (\ref{trcq}).
%%% Simulation part
\section{Performance Evaluation Results}\label{sec:simulation}
In this section, we present computer simulation results for the performance evaluation of the proposed self-supervised hybrid deep learning method. As mentioned in \ref{sec:format}, we focus on the case with single user, and parameters of the two datasets in our simulation are the same as in table \ref{systemparameter}. We assume that WAIR-D\_n refers to scenario-2 and case $n$ of WAIR-D, and DeepMIMO\_n refers to the $n\text{-}th$ 1000 users between row 1000 and row 1300 in O1\_60 of DeepMIMO. Thus, the training dataset would be mixture of WAIR-D\_1 and DeepMIMO\_1, adding up to 1000 users in each training. And the test dataset is chosen from WAIR-D\_2 to WAIR-D\_5 and DeepMIMO\_2 to DeepMIMO\_5. The model is trained through 10000 epochs using TensorFlow 2.8 in Ubuntu 20.04 equipped with an EPYC 75F3 CPU and a RTX 3090 GPU with learning rate being initialized at 0.001 and the batch size being 256. First, we will show the robustness of our proposed scheme among two different datasets, then we will evaluate the influence of the proportion and compare it with the theoretical analysis.
\begin{figure}[h]\vspace{-4mm}
	\begin{center}
		\centerline{\includegraphics[width=0.35\textwidth]{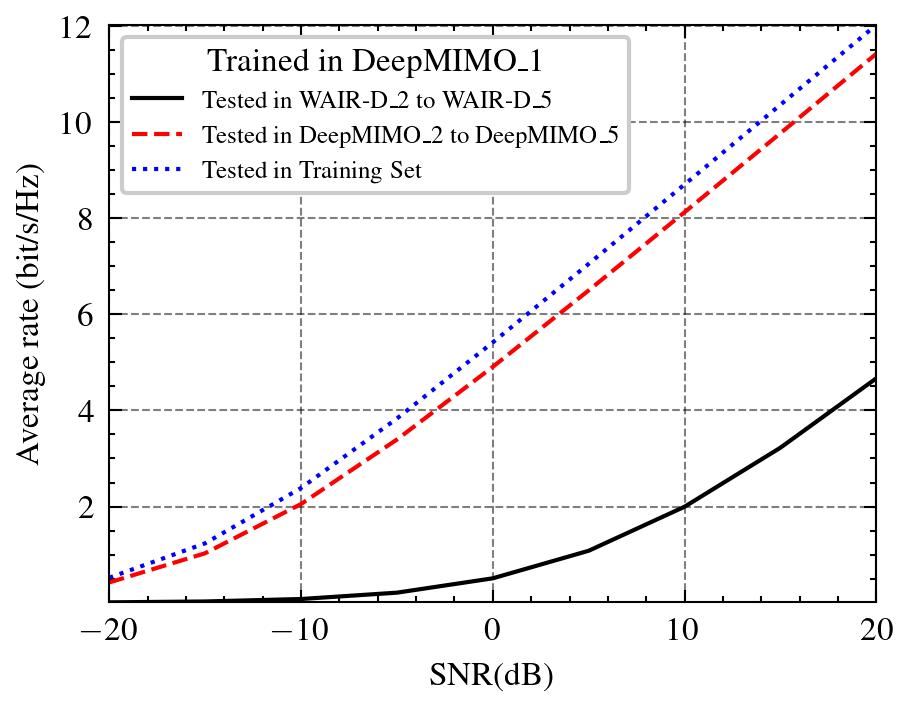} } \vspace{-0mm}
		\caption{Only trained in DeepMIMO\_1.}
		\label{fig:trainindeepmimo}
      \vspace{-2mm}
	\end{center}
\end{figure}\vspace{-4mm}
\begin{table}[h!] \vspace{-4mm} 
\centering
\caption{SYSTEM PARAMETERS}
\label{systemparameter}
\begin{tabular}{||c c c||}
 \hline
 Parameters & DeepMIMO & WAIR-D  \\ [1.0ex]
 \hline\hline
 AP & BS3  & BS0 \\
 Users &  \multicolumn{2}{c|}{1000}  \\
 Carriar Frequency & \multicolumn{2}{c|}{60GHz}  \\
 Radio Chains & \multicolumn{2}{c|}{1}  \\
 Antenna Number & \multicolumn{2}{c|}{64} \\
 Antenna Spacing & \multicolumn{2}{c|}{$ \lambda/2$}  \\
 Bandwidth & \multicolumn{2}{c|}{50 MHz} \\
 Path Number & \multicolumn{2}{c|}{5}  \\
 Signal Power & \multicolumn{2}{c|}{1dbm}\\
 PNR & \multicolumn{2}{c|}{20dB} \\[1.0ex]
 \hline
\end{tabular}
\end{table} \vspace{-2mm}
 
\begin{figure}[h]\vspace{-4mm}
	\begin{center}
		\centerline{\includegraphics[width=0.33\textwidth]{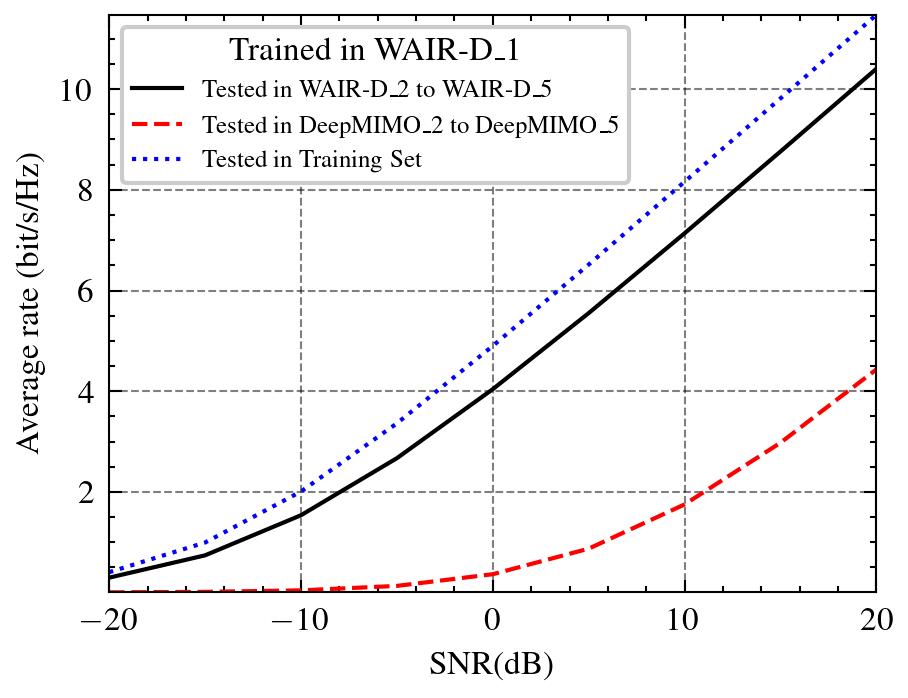} } \vspace{-0mm}
		\caption{Only trained in WAIR-D\_1.}
		\label{fig:traininwaird}
      \vspace{-4mm}
	\end{center}
\end{figure}

Fig$.$~\ref{fig:trainindeepmimo} and Fig$.$~\ref{fig:traininwaird} show that the performance of the beamforming scheme is trained with only one kind of dataset, but are tested in two different datasets respectively.  It shows that two systems with different training datasets have a similar system capacity under the same scenario setting. However, results also show that the beamforming model only works well when we train and test it in the same kind of dataset, which implies the beamforming does not have the robustness to adapt the change of environment.

\begin{figure}[h]\vspace{-2mm}
	\begin{center}
		\centerline{\includegraphics[width=0.33\textwidth]{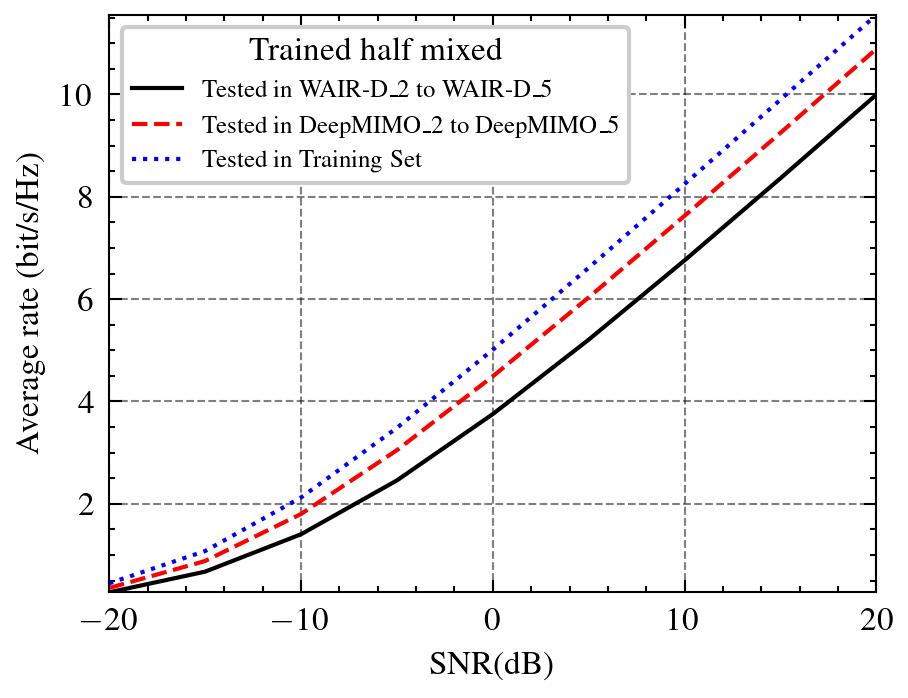} } \vspace{-0mm}
		\caption{Trained half mixed.}
		\label{fig:trainhalfmixed}
      \vspace{-4mm}
	\end{center}
\end{figure}

In Fig$.$~\ref{fig:trainhalfmixed}, we evaluate the performance of the proposed hybrid training technique, in which the two datasets are mixed equally while keeping overall number of users unchanged. It is surprising that the test result in two different datasets do not show significant performance loss as shown in Fig$.$~\ref{fig:trainindeepmimo} and Fig$.$~\ref{fig:traininwaird}, i.e., all test cases achieving similar maximum rate. Therefore, it is concluded that our proposed self-supervised hybrid deep learning scheme has the strong robustness for different datasets. Moreover, this scheme can be extended to more general scenario, i.e., there are many kinds of datasets, then, we can just need to get the training data from these datasets in a certain proportion to feed to the neural network.
\vspace{-4mm}

\begin{figure}[h]\vspace{-2mm}
	\begin{center}
		\centerline{\includegraphics[width=0.33\textwidth]{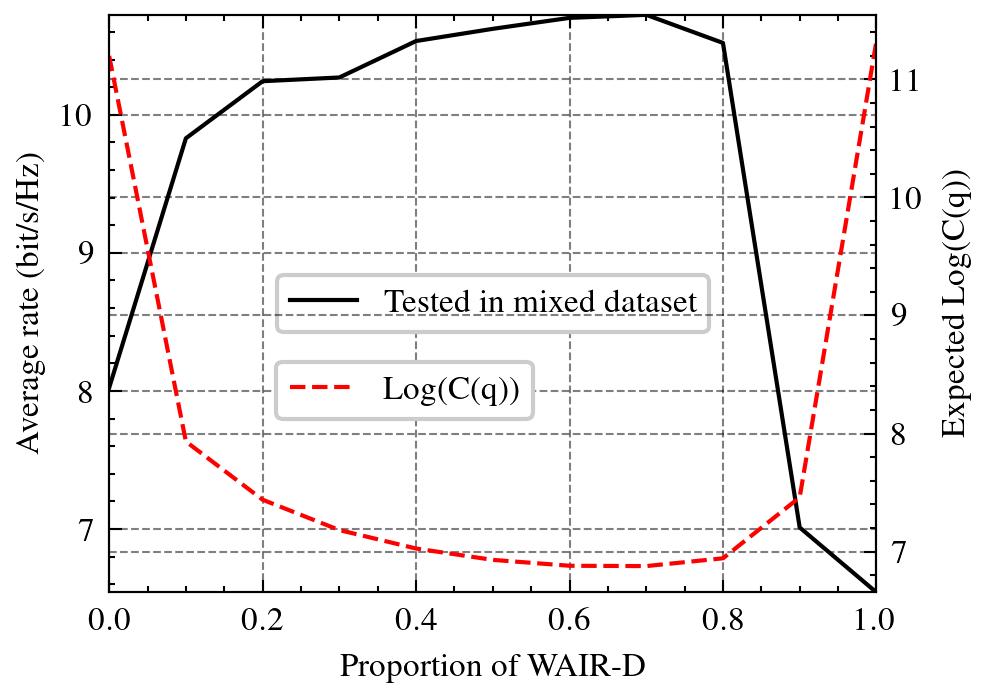} } \vspace{-0mm}
		\caption{Impact of proportion of WAIR-D when evaluated with half mixed dataset and log(C(q)).}
		\label{fig:proportion}
      \vspace{-4mm}
	\end{center}
\end{figure}

In Fig$.$~\ref{fig:proportion}, the impact of proportion $\mathbf q$ on the performance of the proposed hybrid training is evaluated, and compared with the theoretical calculation $\log(C(\mathbf q))$. Specifically, when we constrain the training samples number to 1000 from these two datasets, but increase the proportion of WAIR-D\_1 from 0 to 1 with stepping 0.1. It is found that when the proportion of WAIR-D reaches 70\%, the model outputs the highest average rate, which means that the proposed scheme can work at the best efficiency in both scenarios if we get 70\% data from WAIR-D and 30\% data from DeepMIMO. Similarly, we plot the curve of $\log(C(\mathbf q))$ through the theoretical (\ref{trcq}) and (\ref{trlambda}) function.  It can be seen that it is a U-shaped curve and reaches the minimal value at 70\% for the proportion of WAIR-D, which reversely matches the average rate performance very well. In addition, Fig$.$~\ref{fig:proportion} also shows that too small or too large data proportion will only make the performance unbalanced and biased. This means that we only need to minimize $C(\mathbf q)$ to obtain the best proportion from different training datasets, which makes sure that the trained model has the best robustness performance all datasets.
  \vspace{-2mm}
\section{Conclusion}\label{sec:conclusion}
In this paper, we have proposed a robust beamforming method for mmWave communications utilizing self-supervised hybrid deep learning. By a great deal of simulation experiments, we found that our proposed scheme not only can keep robust inside a dataset, but also it can work well in different datasets. Moreover, a theoretical analysis was also presented to explain the effectiveness, which also fitted well with our experiment results. Moreover, our proposed self-supervised beamforming scheme could be extended to more kinds of datasets, and also has strong robustness. However, more interesting topics such as how to design the best network architecture and the exact function of each layer is still under exploration.
  \vspace{-4mm}
\bibliographystyle{IEEEtran}
\bibliography{Robust_Beamforming_HDL}
\vspace{12pt}

\end{document}